\documentclass[]{pasj01}
\draft

\begin{document} 
\Received{}%{yyyy/mm/dd}
\Accepted{}%{yyyy/mm/dd}
%\Published{yyyy/mm/dd}

\def\Herschel{{\it Herschel}}
\def\Spitzer{{\it Spitzer}}
\def\WISE{{\it WISE}}
\def\AKARI{{\it AKARI}}

\def\um{$\mu \rm{m}$}

\def\Xco{$X_{\rm CO}$}
\def\Xunit{$\rm{cm}^{-2}$ $\rm{K}^{-1}$ $\rm{km}^{-1}$ $\rm{s}$}
\def\cmcm{$\rm{cm}^{-2}$}
\def\cmcmcm{$\rm{cm}^{-3}$}
\def\kms{$\rm{km}$ $\rm{s}^{-1}$}
\def\vlsr{$v_{\rm LSR}$}
\def\degree{$^{\circ}$}
\def\Lsun{$L_{\solar}$}
\def\Msun{$M_{\solar}$}
\def\Msunyr{$M_{\solar}$ $\rm{yr}^{-1}$ }

\def\NH{$N(\rm{H_{2}})$}

\def\HII{H \emissiontype{II}}
\def\OIII{O \emissiontype{III}}
\def\SII{S \emissiontype{II}}

\def\lb{($l$, $b$)}
\def\lbeq{($l$, $b$)$=$}
\def\lbsim{($l$, $b$)$\sim$}
\def\radec{($\alpha_{\rm J2000}$, $\delta_{\rm J2000}$)}
\def\radeceq{($\alpha_{\rm J2000}$, $\delta_{\rm J2000}$)$=$}
\def\radecsim{($\alpha_{\rm J2000}$, $\delta_{\rm J2000}$)$\sim$}

\def\COa{\atom{C}{}{12}\atom{O}{}{}}
\def\COb{\atom{C}{}{13}\atom{O}{}{}}
\def\COc{\atom{C}{}{}\atom{O}{}{18}}

\def\Jeq{{\it J}$=$}
\def\Ja{({\it J}$=1$--$0$)}
\def\Jb{({\it J}$=2$--$1$)}
\def\Jc{({\it J}$=3$--$2$)}

\title{A new view of the giant molecular cloud M16 (Eagle Nebula) in \COa \ \Jeq 1--0 and 2--1 transitions with NANTEN2}

%%% begin:list of authors
% Do NOT capitalize all letters in "textsc".

\author{Atsushi \textsc{nishimura}\altaffilmark{1*}}
\author{Jean \textsc{costes}\altaffilmark{1}}
\author{Tetsuta \textsc{inaba}\altaffilmark{1}}
\author{Kengo \textsc{tachihara}\altaffilmark{1}}
\author{Yusuke \textsc{hattori}\altaffilmark{1}}
\author{Mikito \textsc{kohno}\altaffilmark{1}}
\author{Akio \textsc{ohama}\altaffilmark{1}}
\author{Kazufumi \textsc{torii}\altaffilmark{2}}
\author{Hidetoshi \textsc{Sano}\altaffilmark{1,3}}
\author{Hiroaki \textsc{yamamoto}\altaffilmark{1}}
\author{Yutaka \textsc{hasegawa}\altaffilmark{4}}
\author{Kimihiro \textsc{kimura}\altaffilmark{4}}
\author{Hideo \textsc{ogawa}\altaffilmark{4}}
\author{Yasuo \textsc{fukui}\altaffilmark{1,3}}

\altaffiltext{1}{Department of Physics, Nagoya University, Furo-cho, Chikusa-ku, Nagoya, Aichi 464-8602, Japan}
\altaffiltext{2}{Nobeyama Radio Observatory, National Astronomical Observatory of Japan (NAOJ), National Institutes of Natural Sciences (NINS), 462-2 Nobeyama, Minamimaki, Minamisaku, Nagano 384-1305, Japan}
\altaffiltext{3}{Institute for Advanced Research, Nagoya University, Chikusa-ku, Nagoya, Aichi 464-8601, Japan}
\altaffiltext{4}{Department of Physical Science, Graduate School of Science, Osaka Prefecture University, 1-1 Gakuen-cho, Naka-ku, Sakai, Osaka 599-8531, Japan}

\email{nishimura@a.phys.nagoya-u.ac.jp}
%%% end:list of authors

%% `\KeyWords{}' always has to be placed before `\maketitle'.
 %Do NOT move this preamble from here!

\maketitle
\begin{abstract} % max 300 words.

M16, the Eagle Nebula, is an outstanding \HII \ region where extensive high-mass star formation is taking place in the Sagittarius Arm, and hosts the remarkable “pillars” observed with HST. 
We made new CO observations of the region in the \COa \ \Jeq 1--0 \ and \Jeq 2--1 transitions with NANTEN2. 
These observations revealed for the first time that a giant molecular cloud of $\sim 1.3 \times 10^5$ \Msun \ is associated with M16, which is elongated vertically to the Galactic plane over 35 pc at a distance of 1.8 kpc. 
We found a cavity of the molecular gas of $\sim 10$ pc diameter toward the heart of M16 at \lbeq (16.95\degree, 0.85\degree), where more than 10 O-type stars and $\sim 400$ stars are associated, in addition to a close-by molecular cavity toward a Spitzer bubble N19 at \lbeq (17.06\degree, 1.0\degree). 
We found three velocity components which show spatially complementary distribution in the entire M16 giant molecular cloud (GMC) including NGC6611 and N19, suggesting collisional interaction between them. 
Based on the above results we frame a hypothesis that collision between the red-shifted and blue-shifted components at a relative of $\sim 10$ \kms \ triggered formation of the O-type stars in the M16 GMC in the last 1-2 Myr. 
The collision is two fold in the sense that one of the collisional interactions is major toward the M16 cluster and the other toward N19 with a RCW120 type; the former triggered most of the O star formation with almost full ionization of the parent gas, and the latter an O star formation in N19. 

\end{abstract}
\section{Introduction}

The high mass stars, O- and early B-type stars, are most energetic in the Galactic disk and dominate dynamics of the interstellar medium and influence evolution of galaxies. 
It is therefore crucial to understand the mechanism of high-mass star formation for a comprehensive understanding of galaxy evolution. 
Considerable efforts have been devoted toward this issue and are reviewed by \citet{zin07}.

O/early B stars ionize \HII \ regions which are often associated with molecular gas, a remnant of the placental material. 
\HII \ regions are therefore a primary target to study the mechanisms and initial conditions for high-mass star formation based on observations of stars and molecular gas. 
%Another potential site of promising high-mass stars formation, infrared dark clouds, show often no strong evidence of real O star formation (REF.). 
A difficulty in tracing star formation in \HII \ regions is that the ionization affects significantly the natal gas, which disperses the gas quickly after star formation. 
In order to minimize influence due to the ionization very young star formation is to be chosen for a molecular study, and such samples have to be large for good statistics. 
Recent studies of molecular clouds in Orion, Vela, and the Sagittarius Arm are part of such efforts, which include M42/M43, NGC2023/NGC2024, NGC2068/NGC2071, RCW38, M20, RCW120 etc.; these \HII \ regions show very young star formation with an age of $\sim 10^5$ yr, giving us hints on the initial cloud conditions and distributions prior to O star formation (e.g., \cite{fuk17b, oha18a, tsu18, fuk16, tor11, tor17, tor15}). 

The distance scale affected by the ionization is given as $v_i \times$ age, where $v_i \sim 5$ \kms \ is the velocity of the ionization front (e.g., \cite{spi68}). 
An age of $10^5$ yr yields 0.5 pc as the distance, which is smaller than the typical star forming cloud having a size of a few to $\sim 10$ pc. 
In the regions having age of more than a few Myr, molecular gas within a 10 pc radius becomes significantly ionized if there are ten O stars whose age is $\sim$2 Myr as in Westerlund2, NGC3603, and [DBS2003]179 (\cite{fur09, oha10, fuk14}; Fukui et al. 2018). 
The natal gas directly related to star formation is already lost, whereas it is possible that the rest of the gas outside 10 pc still holds a hint on the cloud initial conditions and interactions. 

\begin{figure}[t]
 \begin{center}
  \includegraphics[width=\textwidth]{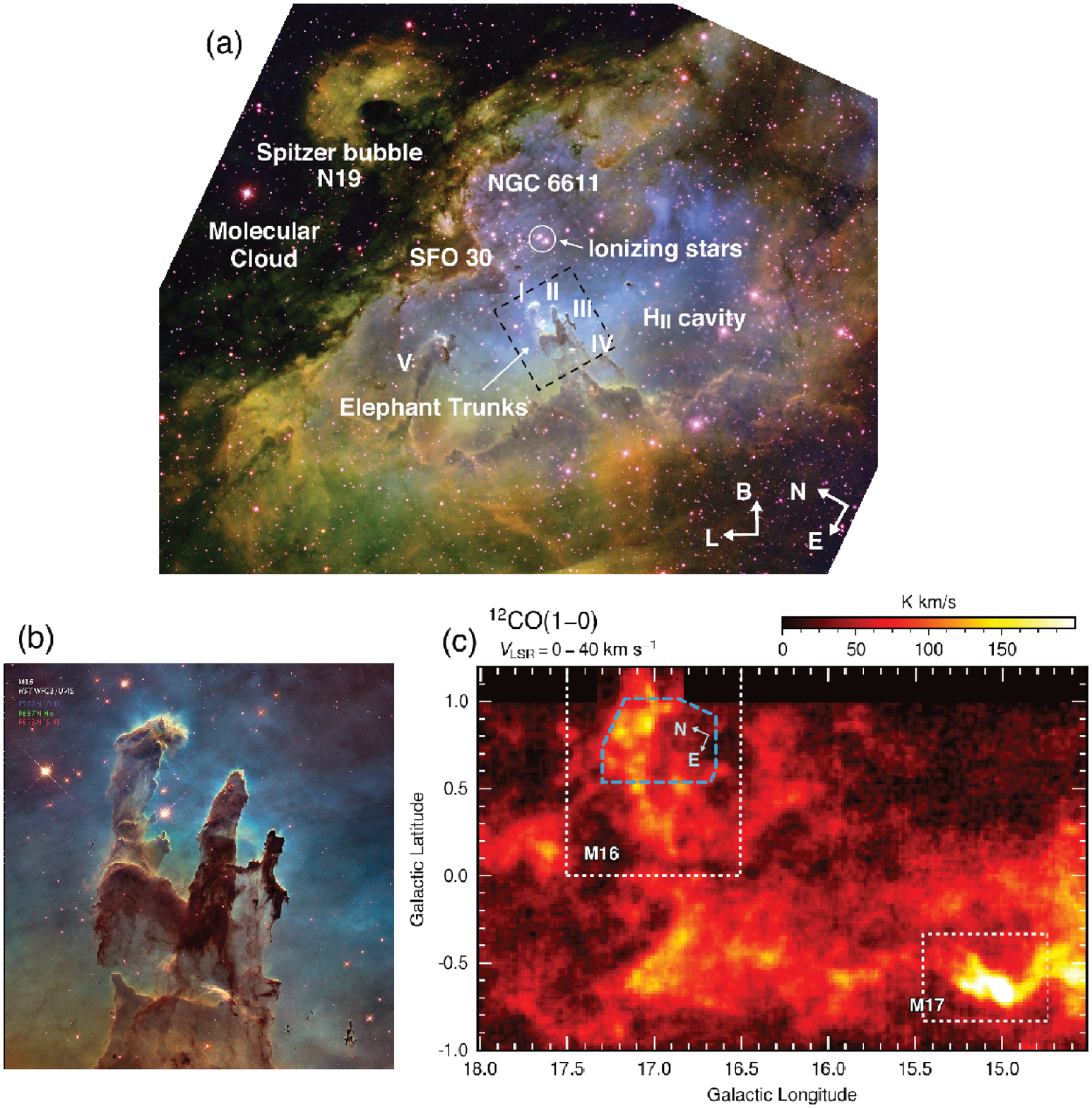} 
 \end{center}
\caption{
(a) Optical composite image of the Eagle Nebula taken with the 0.9 m telescope at the Kitt Peak Observatory with the NOAO Mosaic CCD camera (Credit: T.A.Rector, B.A.Wolpa and NOAO).
Green, blue and red show H$\alpha$, [\OIII] and [\SII]. 
The pillars in Elephant Trunks are labeled from I to V.
The Spitzer bubble N19 \citep{chu06}, the bright rimed cloud SFO30 \citep{sug91} and other famous objects \citep{oli08} are labeled in the figure.
Black dashed lines indicate the region shown in Figure 1b.
(b) Optical multicolor images of the Elephant Trunks taken with Hubble Space Telescope with WFC3/UVIS \citep{lev15}. 
Green, blue and red show H$\alpha$, [\OIII] and [\SII]. 
(c) Large-scale \COa \ \Jeq 1--0 distributions toward the Sagittarius-Carina Arm taken with the NANTEN2 telescope. 
The intensity is integrated over 0 to 40 \kms.
White dotted boxes indicate the area of M16 and M17 GMCs.
Blue dashed lines indicate the region shown in Figure 1a.
}
\label{fig:1}
\end{figure}

It is noteworthy that recent observations suggest a possibility that cloud-cloud collision may be triggering O/early B star formation in \HII \ regions. 
%They include superstar clusters or isolated O stars; e.g., Westerlund2 \citep{fur09, oha10}, NGC3603 \citep{fuk14}, RCW38 \citep{fuk16}, M20, RCW120 \citep{tor15}. 
The M16, the Eagle Nebula, in the Sagittarius Arm, is one of the best-known high-mass star formation sites (for review, see \cite{oli08}; Figure \ref{fig:1}a) at a distance of 1.8 kpc \citep{bon06, duf06, gua07} which consists of an open cluster NGC6611, an \HII \ region known as Sh2-49 \citep{sha59}, Gum83 \citep{gum55}, RCW165 \citep{rod60} or W37 \citep{wes58}, some bright rimmed clouds (e.g., \cite{sug91}), and molecular clouds \citep{dam01}.
NGC6611 consists of 52 OB stars \citep{eva05} including O4 star HD168076.
The age and total mass of the NGC6611 population are estimated to be 1.3$\pm$0.3 Myr \citep{bon06} and  $\sim 2.5 \times 10^{4}$ \Msun \ \citep{wol07}, respectively.
The Initial Mass Function (IMF) in the cluster is measured to be similar to a Salpeter IMF with a slope of $-1.52$ for the entire cluster ($R < 6.5$ pc), whereas relatively flat with a slope of $-0.62$ for the cluster core ($R < 0.7$ pc) \citep{bon06}.
M16 is also remarkable because it exhibits spectacular pillars (also called “elephant trunks”; Figure \ref{fig:1}b), a remnant of the placental molecular gas being ionized by the formed O stars (e.g., \cite{hes96, lev15}). 
The dense molecular cores, known as evaporating gaseous globules (EGGs), are observed at the surface of the pillars \citep{hes96}, whereas association with a massive YSO is rare \citep{mcc02}, and there were no strong evidence for triggered star formation \citep{hes05, ind07}.
The associated molecular material, most likely the parent material forming the stars, may be imagined indirectly from optical obscurations in the north and south of the Eagle Nebula, whereas direct mm/sub-mm observations of molecular gas were not made in a large scale, and a full picture of the cloud did not emerge so far. 
We carried out new CO \Jeq 1--0 and 2--1 observations with NANTEN2 toward M16 and have revealed the whole cloud and its surroundings for the first time at angular resolutions suitable to study star formation and cloud interaction with an \HII \ region. 
The present paper gives the first results of these observations. 
The paper is organized as follows; Section 2 provides description of CO observations, and Section 3 presents the CO results. 
Section 4 presents a cloud-cloud collision model and Section 5 concludes the paper.

\section{Observations}

\subsection{\COa \ \Jeq 1--0 with NANTEN2 telescope}

The observations of \COa \ \Jeq 1--0 emission line at 115 GHz were performed by the NANTEN2 4-m submillimeter telescope of Nagoya University \citep{miz04} and were conducted from January 2012 to January 2013.
The front end was a 4 K cooled superconductor-insulator-superconductor (SIS) mixer receiver.
The double-sideband (DSB) system noise temperature was typically measured to be 110 K toward the zenith including the atmosphere.
The back end was a digital Fourier transform spectrometer (DFS) with 16,384 channels of 1 GHz bandwidth, corresponding to a velocity coverage and resolution of 2,600 \kms \ and 0.16 \kms, respectively.
The on-the-fly (OTF) mapping mode was used for the observations, and each observation was carried on a submap which has an area of one deg$^2$.
The pointing accuracy was checked every 3 hours and achieved within an offset of 25$''$.
The absolute intensity was calibrated by observing IRAS 16293-2422.
After initial data reductions, the final beam size (FWHM) is 180$''$, and the typical noise level is 0.42 K in $T_{\rm mb}$ scale.

\subsection{\COa \ and \COb \ \Jeq 2--1 with NANTEN2 telescope}

The observations of \COa \ and \COb \ \Jeq 2--1 lines were also carried out with the NANTEN2 telescope in November 2015 to obtain improved angular resolution data.
The area of 0.2\degree $\times$ 0.2\degree \ in the center of M16 GMC were observed with an angular resolution of 90$''$.
A 230 GHz band SIS receiver was used in DSB mode with a typical system noise temperature 200 K during observations.
The back end was DFS with a velocity coverage and resolution of 1,300 \kms \ and 0.08 \kms, respectively.
The on-the-fly (OTF) mapping mode was used for the observations.
The pointing accuracy was checked every 3 hours and achieved within an offset of 10$''$.
The absolute intensity was calibrated by observing M17 SW.
After initial data reductions, the final beam size (FWHM) is 90$''$, and the typical noise level is 1.8 K in $T_{\rm mb}$ scale.

\section{Results}

Figure \ref{fig:1}c shows large-scale integrated intensity distribution of the \COa \ \Jeq 1--0 emission in a 3.5\degree $\times$ 2\degree \ region of the M16 and M17 giant molecular clouds (GMCs) obtained with NANTEN2 telescope.
M16 is away from the Galactic plane peaked at \lbeq (17.03\degree, 0.87\degree) and M17 at \lbeq (15.00\degree, $-$0.67\degree). 
There are several less bright clouds between the two GMCs.  

\begin{figure}[t]
 \begin{center}
  \includegraphics[width=\textwidth]{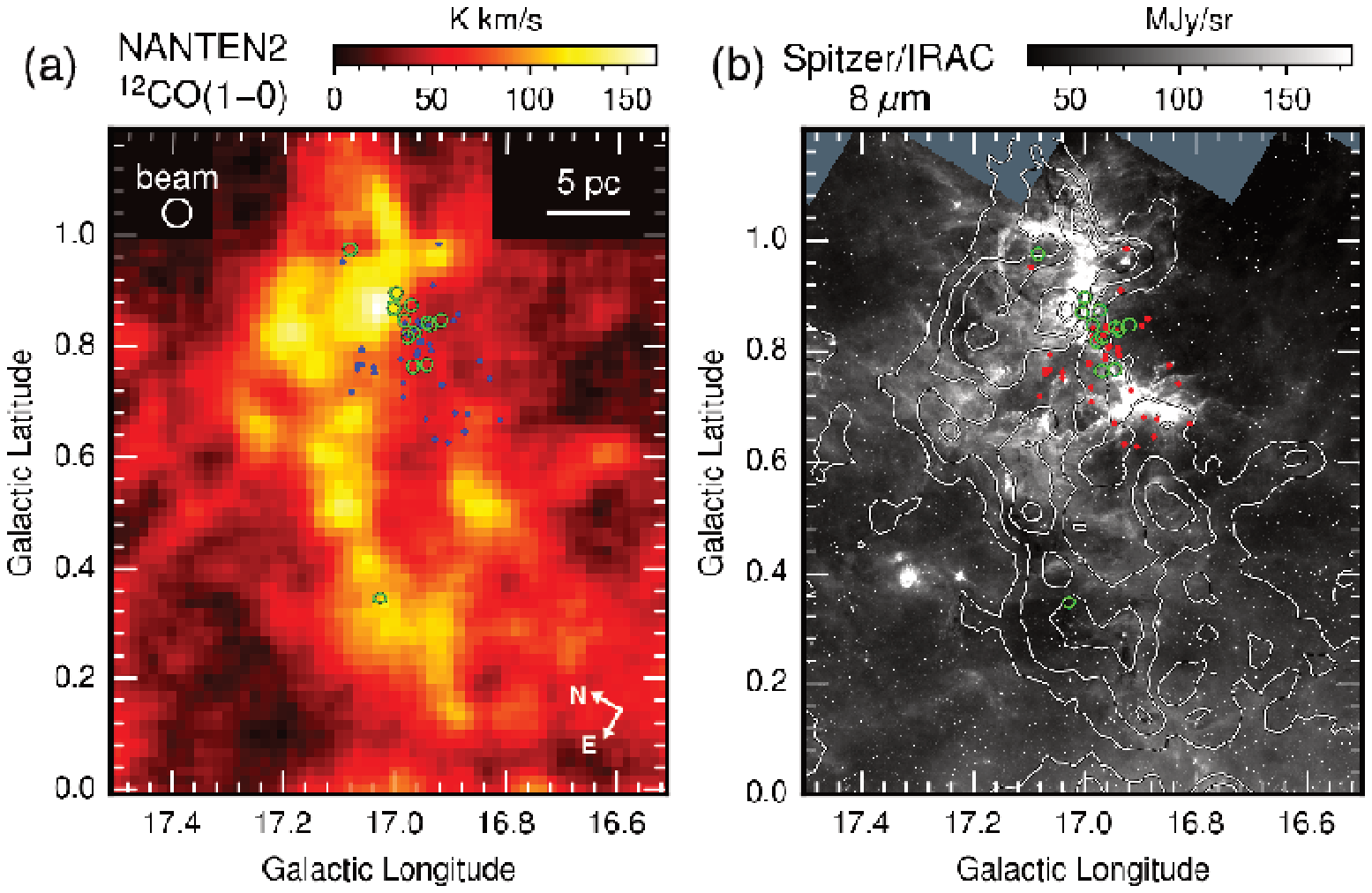} 
 \end{center}
\caption{
(a) An integrated intensity map of \COa \ \Jeq 1--0 toward M16. 
The velocity range used for the integration is 0.0 \kms \ to 40.0 \kms.
Green open circles and blue dots indicate O and B stars \citep{eva05}, respectively.
(b) Distribution of the \Spitzer \ 8 \um \ image superposed on the contours of the integrated intensity of \COa \ \Jeq 1--0 shown in Figure \ref{fig:2}a.
The contours are plotted at every 25 K \kms \ from 55 K \kms.
Green open circles and red dots indicate O and B stars, respectively.
}
\label{fig:2}
\end{figure}

\begin{figure}[t]
 \begin{center}
  \includegraphics[width=\textwidth]{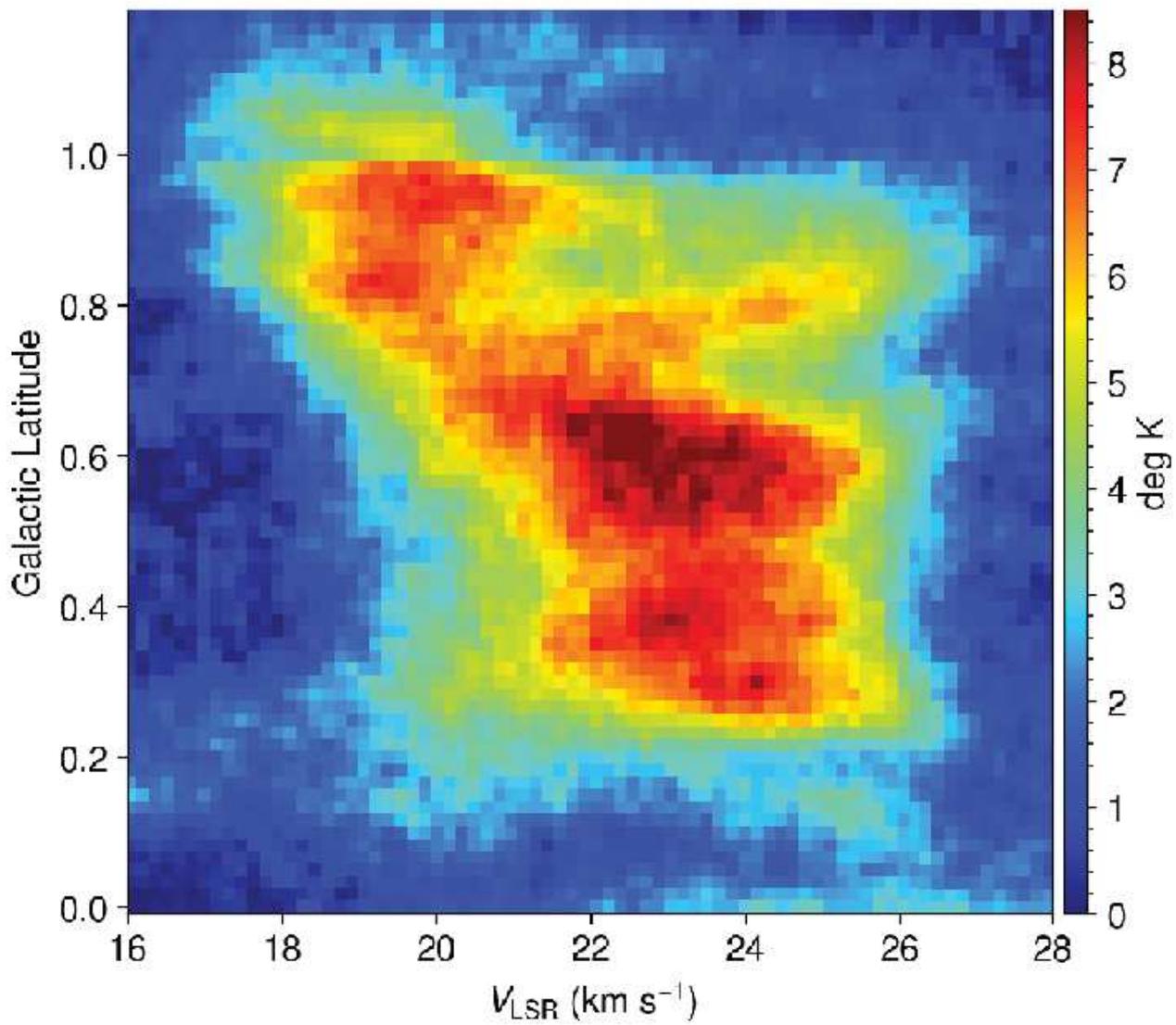} 
 \end{center}
\caption{
Latitude-velocity ($b$-$v$) diagram of the M16 GMC for the emission of \COa \ \Jeq 1--0.
Spectra in the longitude range between $l=$17.5\degree \ and 16.5 \degree \ are used to produce the diagram.
}
\label{fig:3}
\end{figure}

Figure \ref{fig:2}a shows integrated intensity distribution of the \COa \ \Jeq 1--0 emission toward M16. 
The GMC has a main ridge vertical to the plane with a length of $\sim$ 35 pc.
The peak column density is estimated to be $1.7 \times 10^{22}$ \cmcm \ using \Xco \ factor of $1.0 \times 10^{20}$ \Xunit \ \citep{oka17}.
The total mass of the M16 is estimated to be $1.3 \times 10^5$ \Msun.
Figure \ref{fig:2}b shows an overlay of the CO contours with the Spitzer infrared image at 8 \um. 
The M16 cluster is located at the south of the main CO peak, where CO emission shows a sharp decrease. 
The size of this CO cavity is $\sim 10$ pc in diameter.
The pillars are located in the cavity at \lbeq (16.94\degree, 0.77\degree). 
We found another small depression of CO intensity at \lbeq (17.1\degree, 1.0\degree), which corresponds to the O9 star W584 \citep{wal61} the exciting source of a Spitzer bubble N19 \citep{chu06} in the north edge of the M16 GMC. 
We found no high-mass star formation signature in the GMC below $b<0.6$\degree \ except for an O8 star HD168504.
Figure \ref{fig:3} shows a latitude-velocity diagram of the M16 GMC. 
There is a distinct difference in velocity of the GMC along the Galactic latitude; the GMC above $b=0.7$\degree \ has two peaks at 19.5 \kms \ and 25 \kms, while that below $b=0.8$\degree \ is singly peaked at 23 \kms. 
So, only the upper side of the GMC is associated with the blue-shifted component.

\begin{figure}[t]
 \begin{center}
  \includegraphics[width=\textwidth]{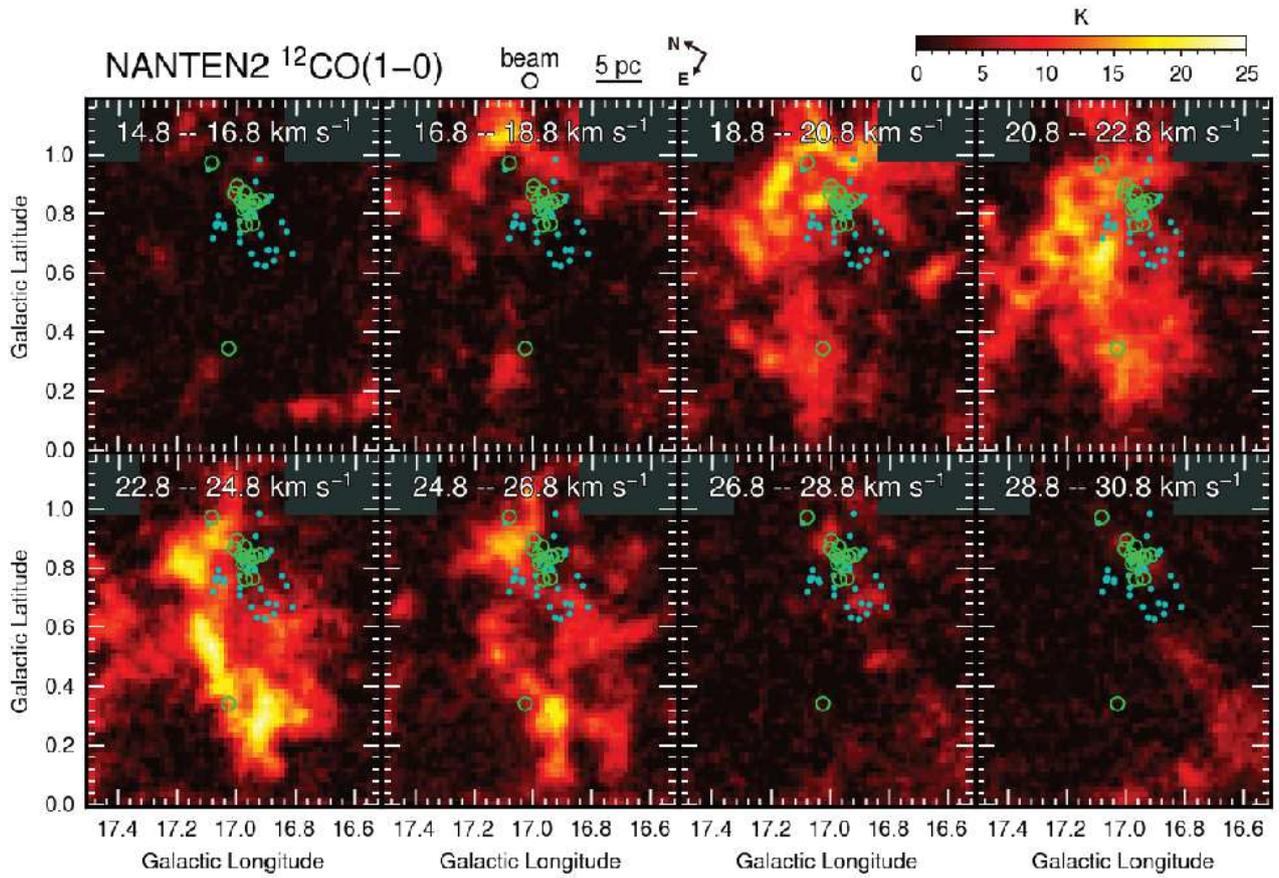} 
 \end{center}
\caption{
Velocity channel map of of the \COa \ \Jeq 1--0 emission at a velocity step of 2 \kms \ toward M16, using the same observed area as Figure \ref{fig:2}.
The velocity range used for averaging are indicated in the top of each panel.
Green open circles and blue dots indicate O and B stars \citep{eva05}, respectively.
}
\label{fig:4}
\end{figure}

Figure \ref{fig:4} shows velocity channel distributions of \COa \ \Jeq 1--0 emission.
The M16 GMC can be divided into three different velocity structures: the blue-shifted component (\vlsr $=$14.8--18.8 \kms), the main component (\vlsr $=$18.8--26.8 \kms), and the red-shifted component (\vlsr $=$26.8--30.8 \kms).
The red-shifted component is mainly distributed in the north side of the cluster and clearly corresponds to the depression at the Spitzer bubble N19.
The main component consists of the main ridge of M16, which is elongated with the Galactic latitude.
The depression of CO intensity around the M16 \HII \ region is in the main component.
The red-shifted component consists of weak emission surrounding the cluster and a diffuse tail-like structure which is distributed toward the south of the cluster.

\begin{figure}[t]
 \begin{center}
  \includegraphics[width=\textwidth]{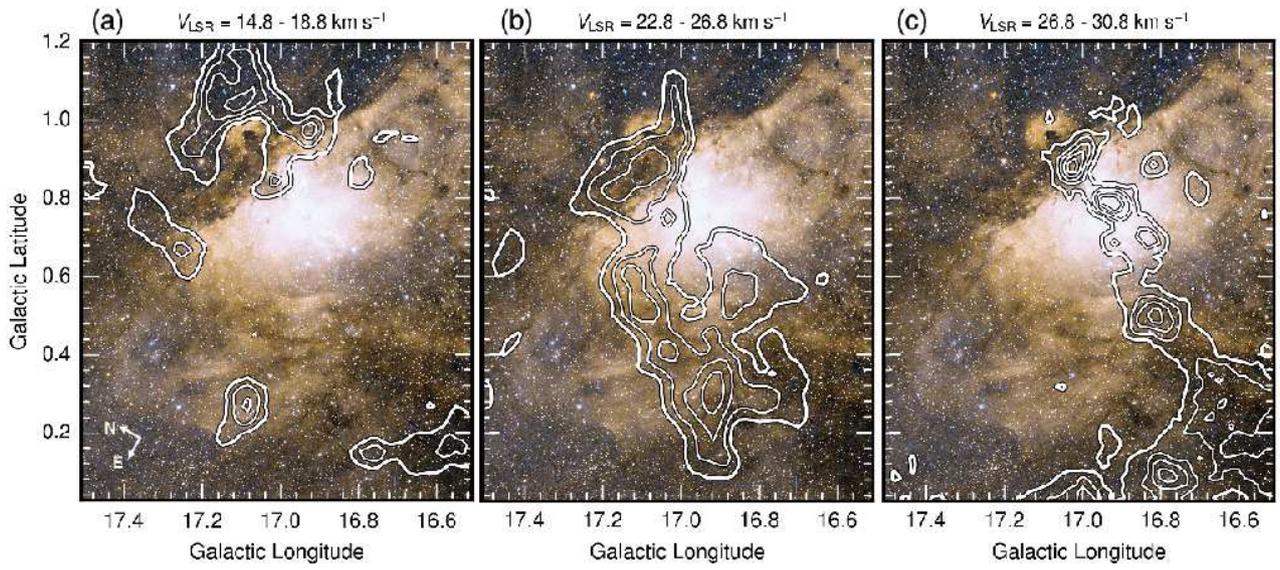} 
 \end{center}
\caption{
Optical images of M16 \HII \ region.
The DSS Red and DSS Blue images are shown in red and blue, respectively.
The averaged image of DSS Red and DSS Blue is used for green.
Contours indicate integrated intensity of \COa \ \Jeq 1--0 emission in the velocity range of (a) 14.8--18.8 \kms \ with contour levels for every 8 K \kms \ from 12 K \kms,
(b) 22.8--26.8 \kms \ with contour levels for every 15 K \kms \ from 25 K \kms, and
(c) 26.8--30.8 \kms \ with contour levels for every 5 K \kms \ from 7 K \kms.
}
\label{fig:5}
\end{figure}

\citet{pou98} observed the Elephant Trunks with \COa \ \Jeq 1--0 line using BIMA interferometer, and found the pillars are associated with molecular gas with a velocity of \vlsr $=$ 20--27 \kms.
They also found a 18--20 \kms \ component and a 27--31 \kms \ component at the position of the pillars (see Figure 3 of \cite{pou98}), but these components have no coincidence with optical dark structures.
Figure \ref{fig:5} shows comparisons of an optical image and velocity structures of the CO emission in a large scale.
The blue-shifted component shows clear anti-correlation with the Spitzer bubble N19 except for the lower side in $b$.
The main component is distributed around the \HII \ region and likely corresponds to the optical dark structure, especially at the north side of the \HII \ region.
The red-shifted component is not coincident with optical structures.
The correspondence between an optical image and CO velocity structures are roughly similar with \citet{pou98}.

\begin{figure}[t]
 \begin{center}
  \includegraphics[width=0.8\textwidth]{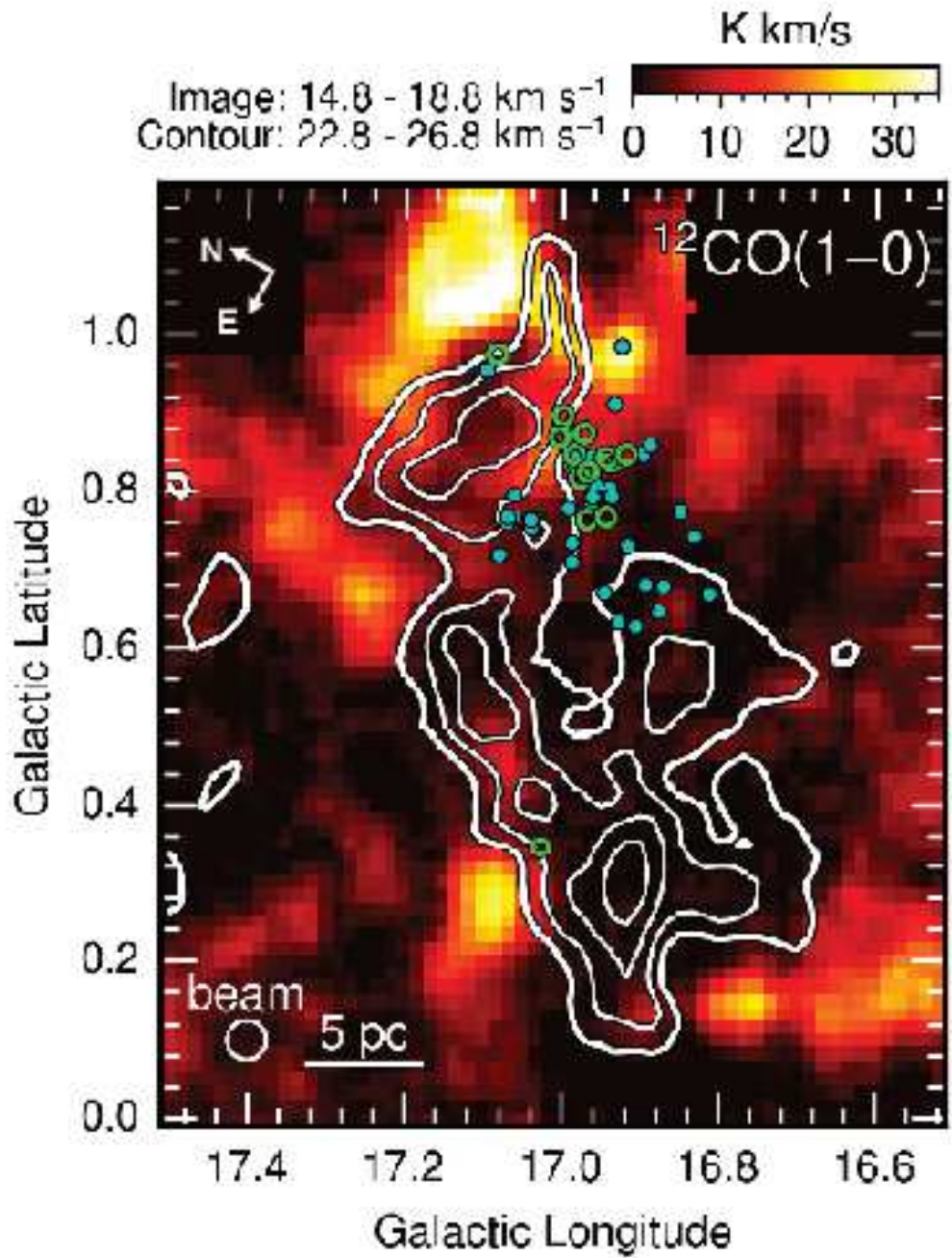} 
 \end{center}
\caption{
Large-scale complementary distributions of the two velocity distributions of \COa \ \Jeq 1--0 emission.
Image indicates integrated intensity of blue-shifted component at the velocity range of 14.8 \kms \ -- 18.8 \kms.
Contours indicate integrated intensity of main component at the velocity range of 22.8 \kms \ -- 26.8 \kms \ with contour levels for every 15 K \kms \ from 25 K \kms.
Green open circles and blue dots indicate O and B stars \citep{eva05}, respectively.
}
\label{fig:6}
\end{figure}

\begin{figure}[t]
 \begin{center}
  \includegraphics[width=\textwidth]{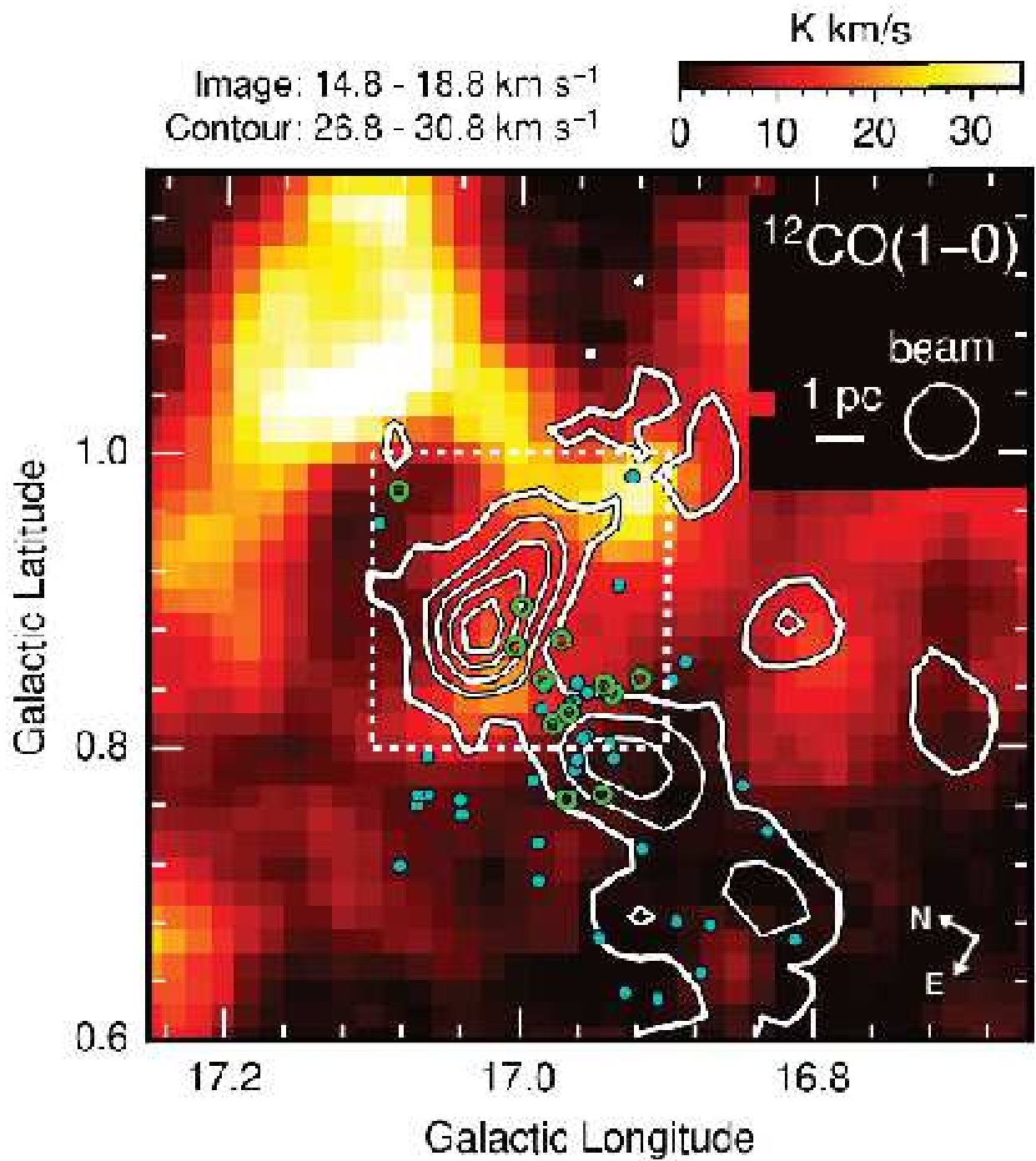} 
 \end{center}
\caption{
The complementary distributions for M16 region.
Image indicates integrated intensity of blue-shifted component at the velocity range of 14.8 \kms \ -- 18.8 \kms.
Contours indicate integrated intensity of red-shifted component at the velocity range of 26.8 \kms \ -- 30.8 \kms \ with contour levels for every 5 K \kms \ from 7 K \kms.
Green open circles and blue dots indicate O and B stars \citep{eva05}, respectively.
White dotted box indicates observation field of \COa \ \Jeq 2--1 shown in Figure \ref{fig:8}.
}
\label{fig:7}
\end{figure}

We find complementary distribution between the blue-shifted component and the main component (Figure \ref{fig:6}).
The blue-shifted component surrounds the main component at the north side of the \HII \ region.
In the lower side in latitude, the blue-shifted component also seems to surround the main component, and an O8 star HD168504 is located on the border of the two components.
The depression likely represents a cavity formed by ionization, whereas the extension of the depression below $b=0.7$\degree \ may not be due to ionization but an intrinsic cloud distribution. 
Figure \ref{fig:7} shows a close-up view of M16 \HII \ region with the blue-shifted component and the red-shifted component, which also show complementary distribution.
The red-shifted component has a peak at \lbeq (17.02\degree, 0.87\degree) on the north side of the \HII \ region, behind the local peak of the blue-shifted component along the direction of the cluster. 
The red-shifted component has a local peak intensity at \lbeq (16.92\degree, 0.78\degree), which corresponds to the intensity depression of the blue-shifted component.

\begin{figure}[t]
 \begin{center}
  \includegraphics[width=\textwidth]{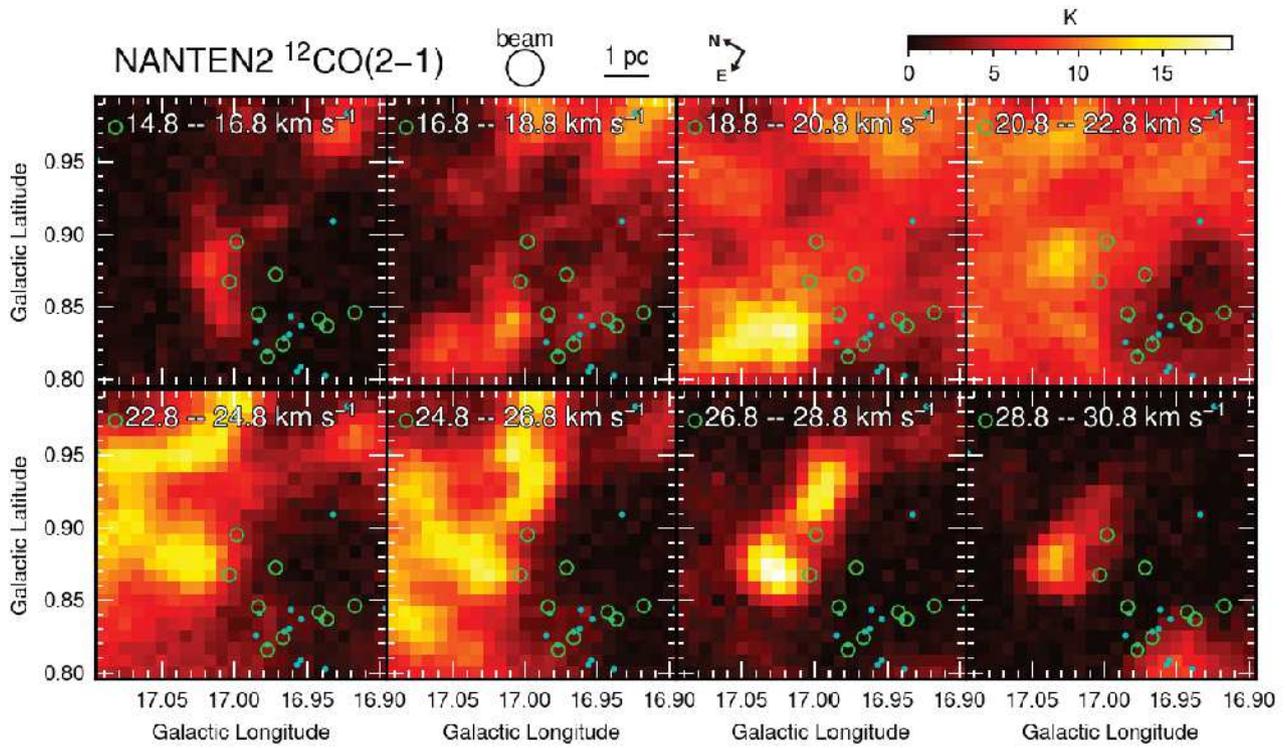} 
 \end{center}
\caption{
Velocity channel map of of the \COa \ \Jeq 2--1 emission at a velocity step of 2 \kms \ toward the center region of M16.
The velocity range used for averaging are indicated in the top of each panel.
Green open circles and blue dots indicate O and B stars \citep{eva05}, respectively.
}
\label{fig:8}
\end{figure}

\begin{figure}[t]
 \begin{center}
  \includegraphics[width=\textwidth]{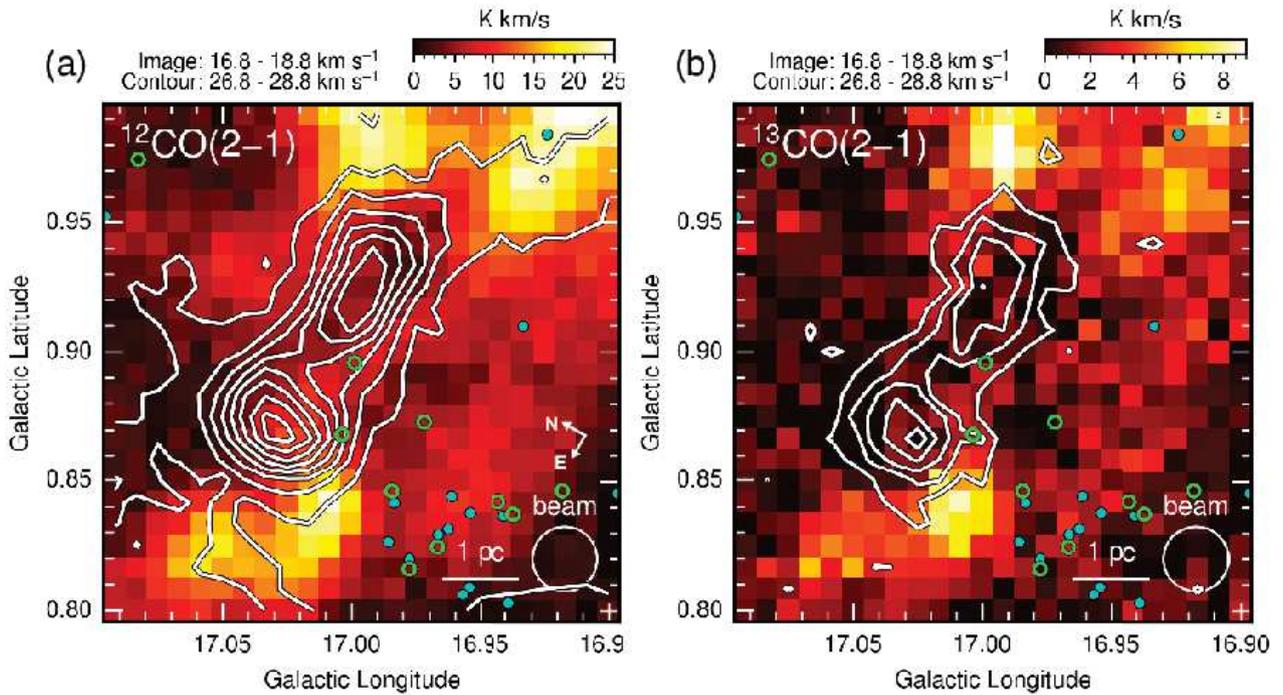} 
 \end{center}
\caption{
The complementary distributions for the center region of M16.
Image and contours indicate blue-shifted component and red-shifted component, respectively, for (a) \COa \ \Jeq 2--1 emission with contour levels for every 4.5 K \kms \ and (b) \COb \ \Jeq 2--1 emission with contour levels for every 2.5 K \kms.
Green open circles and blue dots indicate O and B stars \citep{eva05}, respectively.
}
\label{fig:9}
\end{figure}

\begin{figure}[t]
 \begin{center}
  \includegraphics[width=\textwidth]{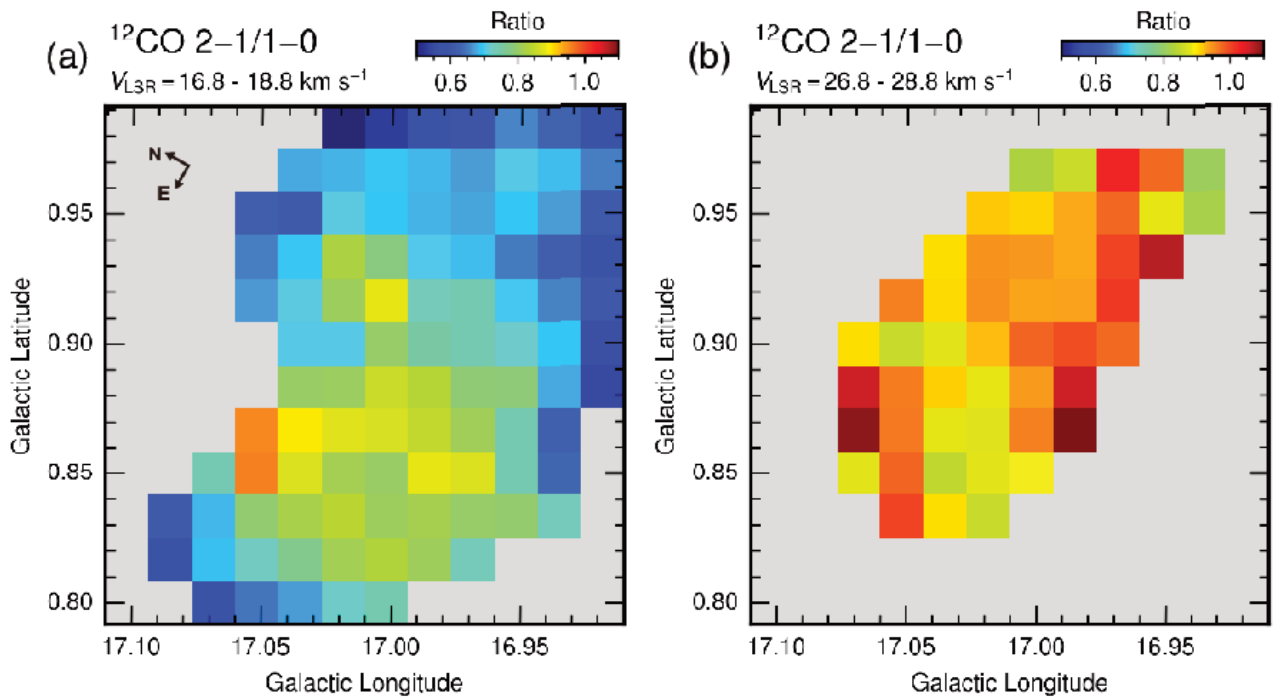} 
 \end{center}
\caption{
Distribution of the \COa \ \Jeq 2--1 / \Jeq 1--0 intensity ratio for (a) blue-shifted component and (b) red-shifted component.
Green open circles and blue dots indicate O and B stars \citep{eva05}, respectively.
}
\label{fig:10}
\end{figure}

Figure \ref{fig:8} shows close-up channel maps for an area of 0.2\degree $\times$0.2\degree centered on M16, and Figure \ref{fig:9} shows complementary distribution of the blue-shifted component and the red-shifted component.
The red-shifted component has a clump with size of 1 pc $\times$ 3 pc which consists of two local peaks.
The blue-shifted component shows four local peaks which is located at \lbeq (17.01\degree, 0.83\degree), (17.05\degree, 0.83\degree), (16.92\degree, 0.98\degree), and (16.99\degree, 0.98\degree).
The red-shifted component is located between the local peaks of the blue-shifted component.
Figure \ref{fig:10} shows distribution of the line intensity ratio between the \Jeq 2--1 and \Jeq 1--0 transitions in the blue-shifted component and the red-shifted component.
The ratio is typically $\sim 0.6$ for molecular clouds without \HII \ region \citep{yod10} and $>0.8$ for GMC with \HII \ region \citep{nis15}.
Both velocity ranges show high ratios above 0.8, supporting their association with the O star cluster.

\section{Discussion}

In the present study we found that the M16 cloud consists of three velocity components with significantly different spatial distributions. 
A careful inspection of the velocity channel distributions in the \COa \ \Jeq 1--0 transition revealed that the blue-shifted component (\vlsr $=$14.8--18.8 \kms) and the main component (\vlsr $=$22.8--26.8 \kms) show complementarity on a large scale.
The complementary distribution in different velocity channels is possible evidence for collision between those clouds (e.g., \cite{fur09, oha10, dob14, shi13, fuk17b, hab92, tak14, mat15}).
The isolated O star W584 in the Spitzer bubble N19, is located on the border of two velocity components, indicating that the stars was possibly formed by cloud-cloud collision following a scheme proposed for a Spitzer bubble RCW120 by \citet{tor15}, where a small cloud collided with the large cloud, created a cavity and compressed a layer in the large cloud.
Since the velocity separation of the two clouds is high ($\sim$8 \kms), cloud collision is able to achieve high mass accretion rate up to 10$^{-3}$--10$^{-4}$ \Msunyr \ \citep{ino13} if they once collided.
HD168504 located on the border of the two velocity components also suggests that the star formation was triggered by the collision of the clouds.

For the small scale, we also found complementary distribution of CO gas around the M16 cluster between the blue-shifted component (\vlsr $=$14.8--18.8 \kms) and the red-shifted component (\vlsr $=$26.8--30.8 \kms).
From detailed comparisons of the CO \Jeq 2--1 distributions (Figure \ref{fig:9}), the two velocity components are likely colliding.
The high intensity ratios ($>0.8$) of the two components indicate their association with the O star cluster that supports colliding scenario.
O stars are located in the area where both blue- and red-shifted components are distributed (Figure \ref{fig:7}).
O stars show elongated distribution in the north--south direction, while the red-shifted component also shows elongated distribution with the north--south direction.
At the north to the cluster, the blue-component remains in molecular state, whereas at the south, most of the blue-shifted component is ionized and observed as the cavity in CO emission.
The radial velocity of the ionized gas is estimated to be $\sim 15.5$ \kms \ \citep{geo70}.
This asymmetric CO distribution along the cluster can be interpreted as a difference of the initial gas distribution and/or an age gradient of the cluster members.
\citet{gua10} found a stellar age gradient in NGC6611, where the southern part of the cluster is older with an age of 2.6 Myr and the northern part in younger with an age of 0.3 Myr.
The radial velocity of NGC6611 member is estimated to be $\sim 15$ \kms \ \citep{bos99, eva05}, which is lower than the velocity of the molecular gas.
This is consistent with the collision scenario because the gas exchanges momentum effectively in the colliding clouds, whereas interaction between stars and gas occurs much less.
In the collision scenario, stars are formed quickly after the collisional event in order of 10$^{5}$ yr, and therefore, stars are expected to keep the initial velocity of the colliding main cloud.
It is impossible to construct a collision picture of the clouds into detail related to NGC6611 formation because the O star cluster heavily ionized surrounding molecular material which contains its initial conditions.
We, however, conclude from the above circumstances, the NGC6611 cluster is possibly formed by cloud-cloud collision.
The velocity separation 10 \kms \ can produce high-mass stars by collision \citep{ino13}.
The collision time scale at the north of the \HII \ region is roughly estimated to be $\sim 3 \times 10^{5}$ yr as the cloud size ($\sim$ 3 pc) divided by velocity separation (10 \kms).
The collision may be still on-going in the GMC, and therefore, it is a lower limit for the collision of the whole star formation region probably triggered formation of NGC6611.
By considering more than 20 cases of O star formation triggered by cloud-cloud collision (e.g., \cite{fuk15, fuk17a, fuk18, oha18a, oha18b, nis18, tsu18} etc. see Section 1), it is a viable scenario that the collision triggered formation of the O stars in M16 and N19 via strong shock compression.

To sum up, the M16 GMC experienced cloud-cloud collision with two collisional spots with O star formation. 
One toward the M16 O star cluster with high column density and the other toward a single O star W584 in N19. 
In the M16 region ionization is significant in dispersing the two parent clouds and only a remnant of the main cloud is seen as the pillars currently. 
The two colliding clouds are still largely surviving in N19 with much less ionization. 
The process has a collision timescale in the order of Myr. 

Formation of N19 is very likely driven by collision, and the molecular circumstance of M16 is consistent with the collision scenario. 
Nonetheless, the present scenario does not exclude another possibility that the GMC collapsed gravitationally without triggering because the cloud dissipation by ionization in M16 hampers to directly watch the cloud-cloud collision toward M16. 
This is a usual limitation in testing a collision scenario under strong ionization. The M16-M17 region has several CO clouds along the Galactic plane (Figure 1). 
They may be potential candidates for future active star formation because the high number density of GMC favors more frequent collision than the average in the Galactic disk. 
Cloud-cloud collision triggering massive cluster in M17 has been also discovered by \citep{nis18}. 
The mean free time of GMC collision is less than 10 Myr within the solar circle (Tachihara et al. 2017), and the M16-M17 region, where the GMC number density is probably enhanced in the Sagittarius Arm, has an even shorter mean free time, favoring collision. 
The vertical distribution at the high Galactic latitude around 1 degree suggests that the M16 GMC was formed by some active event which lifted up molecular gas $\sim 100$ pc above the Galactic plane, possibly being driven by a massive stellar cluster in the last $\sim 10$ Myr. 
The M17 GMC may share a common origin, and the present collision in M16 may have been driven by such an active event.

\section{Conclusions}

We made a large-scale CO observations of the region of M16, the Eagle Nebula, and a Spitzer bubble N19 in the CO $J=$1--0 and $J=$2--1 transitions with NANTEN2. 
A giant molecular cloud of $4 \times 10^5$ \Msun \ is associated with M16, which is elongated vertically to the Galactic plane over 35 pc at a distance of 1.8 kpc. 
The cloud consists of two velocity components in its northern region of $\sim 20$ pc length, and the two components show spatial correlations as complementary distribution suggesting the dynamical interaction. 
We summarize the present work as follows;

\begin{enumerate}
\item
The M16 GMC has a size of 20 pc $\times$ 35 pc in $l$ and $b$, and shows three velocity components of 14.8--18.8 \kms \ (blue-shifted component), 22.8--28.8 \kms \ (main component) and 28.8--30.8 \kms \ (red-shifted component). 
The total mass of the GMC is estimated to be $1.3 \times 10^5$ \Msun from \COa \ \Jeq 1--0 transition by using an \Xco \ factor.

\item
The GMC is associated with 52 OB stars most of which are within the Eagle Nebula, and the O stars are lined up in a straight line of $\sim 10$ pc with a position angle of $\sim 45$ deg. 
We see a clear intensity depression of molecular gas toward the O stars which are likely due to photo ionization; one is toward the heart of the Eagle Nebula at ($l$,$b=$16.95\degree, 0.85\degree) and the other toward the Spitzer bubble N19 ($l$,$b=$17.06\degree, 1.0\degree). 

\item
We find complementary distribution between the blue-shifted and main components toward the entire M16 GMC and between the red-shifted and blue-shifted components toward the center of NGC6611 cluster. 
The molecular distribution in N19 is not fully ionized yet probably because the ultraviolet flux of the exciting star of N19, a O9 star, is not strong enough to ionize the molecular gas in the order of Myr. 
Contrary, the molecular gas which were distributed toward the heart of M16 was almost fully ionized by the 11 O stars, and there remain several small cloudlets at 25 \kms, which are still surviving against the photo ionization and are seen as the pillars resolved with the HST. 
The ionization is due to the enhanced ultraviolet flux by the 11 O stars in M16. 

\item
Based on the above results we hypothesize that collision between the blue-shifted and red-shifted components at a relative of $\sim 10$ \kms \ triggered formation of the O stars in the M16 GMC. 
The collision is two fold in the sense that one of them is toward M16 and the other toward N19. 
The collision timescale is estimated to be $\sim 3 \times 10^5$ yr from the cloud size and velocity separation. 

\end{enumerate}

\begin{ack}

This work was supported by JSPS KAKENHI grant numbers,
15K17607% Torii
.
The authors would like to thank the members of the NANTEN2 group for support on the observation and telescope operation. 
This research made use of astropy, a community-developed core Python package for Astronomy \citep{ast13}, in addition to NumPy and SciPy \citep{wal11}, Matplotlib \citep{hun07} and IPython\citep{per07}.
This research has made use of the VizieR catalogue access tool, CDS,
 Strasbourg, France. The original description of the VizieR service was published in \citet{och00}.
\end{ack}

\end{document}